# Parameter monitoring system of the Dubna Gas-Filled Recoil Separator.

Yu.S.Tsyganov, A.N.Polyakov, A.M. Sukhov.

*Abstract*– **PC-based one-crate monitoring and control system of the Dubna Gas Filled Recoil Separator (DGFRS)** is considered. It is developed for the long-term experiments at the U400 FLNR cyclotron and is aimed at the synthesis of super heavy nuclei in heavy ion induced complete fusion reactions. Parameters related to:
  a) beam and cyclotron;
  b) separator by itself,
  c) detection system,
  d) target and entrance window
are measured and stored in the protocol file of the experiment. Special attention is paid to generating the "alarm" signals and implementing further the appropriate procedures.

## I. INTRODUCTION

During the long term experiments aimed to the synthesis of superheavy elements at the Dubna Gas-filled Recoil separator (DGFRS) the system to measure technological parameters of the experiment and to provide a definite response to some abnormal (alarm) situations is strongly required [1,2].

Usually, a list of parameters/signals includes the following:
- Dipole, quadruples current values measurements as well as setting of alarm thresholds;
- Rotation speeds both entrance window and radioactive target wheels;
- Pressure value in working area of the DGFRS and pentane pressure value in the TOF (time-of-flight module);
- Temperature parameters;
- Beam associated parameters;
- Vacuum parameters;
- Pressure of saturated vapor of pentane in liquid pentane volume
- Photo-diode rotation target output signal amplitude
- ..others.

Manuscript received May 21, 2010. This work was supported in part by the RFBRr Grant No.

All authors are with the Joint Institute for Nuclear Research, 141980 Dubna, Russia. (Telephone: 7-49621-67590, e-mail:tyura@sungns.jinr.ru).

## II. SCHEMATICS OF THE SYSTEM

In the **Fig. 1** the system schematics is shown. Design in brief: CAMAC, one (digital) crate, KK012 controller (modified, see Ref. [2]), program (Windows XP, Borland's Builder C++)

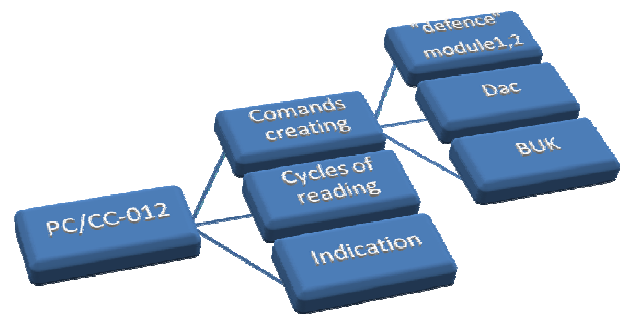

Fig. 1. Schematics of the system

**Main CAMAC modules in the digital apparatuses crate:**
**BUK01** – CAMAC-1M (to create time intervals window to measure parameters, **NF16A0**(3 in parallel, 0.02,02 or 2 s interval) + (three independent outputs) **NF16** (A2/A3, A4/A5, A6/A7)
**BZ01** ("alarm" modules ( 2 mod 2M) each one 8 inputs and 1 output to switch the beam OFF) ; **F24/26( A0-A7)** – off/on "defense" mode, **NF2**(read), **NF10** (to reset alarm's register)
3  16 bit counters KS019
1 ADC PA 01 (8 inp ADC),
2 ADC PA-24K (10 bit, target control, TOF U400(in reserve use)
5 KS022 rate meters (~rotation control, +some others)
1 KV009 1M - DAC → thresholds setting for low Dipole magnet current and Wobbler current

**Gauges**(sensors): **MCS** Baratrons (N=4), vacuum gauges "**Pfiefer**" (N=7, long scale!),
Target (entrance window) - el.Motors(N=2) **"Siemens"** asynchronous AC.
- **Basic dc parameters** measurements: voltage-frequency-code conversions

- Main code **Builder Timers** – 5 (CAMAC, calibration, visualization, imitation of rotation, protocol writing… etc…)

In the **Fig. 2** user interface is shown. Picture (Prt Scr) corresponds to real SHE experiment with $^{48}$Ca ion as a projectile.

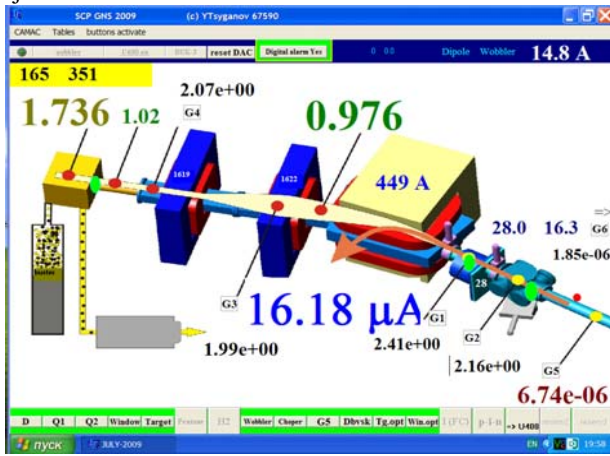

Fig. 2. Schematic of the system interface (three monitors- near the separator, in the control room and near PC).

1) **(examples of values meanings)**:
   a) green field means the parameter under control, if alarm will occur- color will red. Action – beam is turned off;
   b) **16.18**- value (~ add 10%) of projectile beam at the target;
   c) **1.736** (Tor) – pentane pressure;
   d) **28.0 and 16.3** – target and entrance window wheel rotation speed (1/s)// double control: rotating itself+ optical pairs light source- photodiode;
   e) **0.976** – H$_2$ pressure in the separator;
   f) **6.74-06** - cyclotron vacuum value in the point before the separator window;
   g) **449**, **1622, 1619** A – current values in Dipole magnet and in the lenses;
   h) **165    351** at yellow field (left-upper corner) – rates of DAS events (focal plane PIPS + side detector) and TOF camera operation; if **165 351** then it means, that rate of events is under control using low limit ( dblClick in Edit area, ~5 events in ~ 10 s)
   i) **14.8 A** (right-upper corner) – wobbler current
   j) Green button in the left-upper corner: start spectrum measure from additional detector( p-i-n, 8x8 mm$^2$) located close to rotating target in order to estimate it's state
   k) "buster" – pressure of saturated pentane vapor ( in progress)

As examples, graphs for pentane pressure, beam intensity and U-400 $^{48}$Ca ions energy against time are shown in the **Fig. 3a-c**

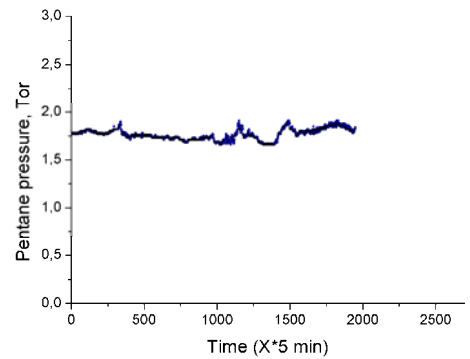

Fig. 3a. The dependence of the pentane pressure in the TOF module against time (no any feed back!).

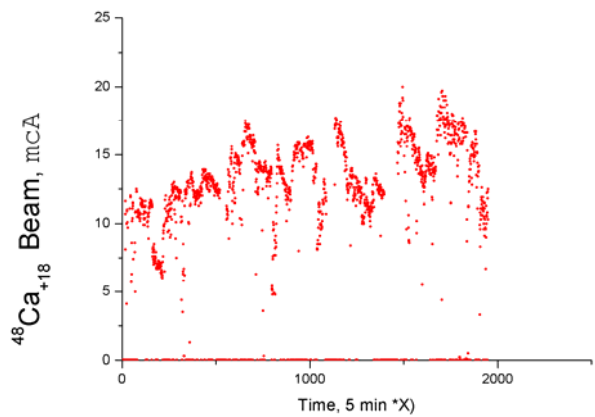

Fig. 3b. The dependence of the bean intensity against time.

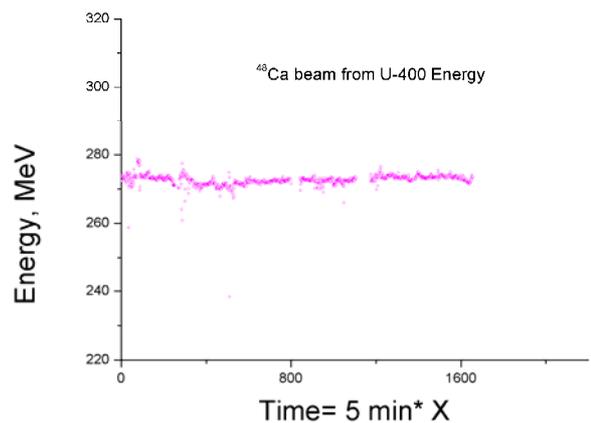

Fig. 3c. Energy against time dependence ( modules to convert pic-Up electrodes signals into spectroscopy codes: ORTEC fast preamps VT 120, ORTEC CFD 584, TAC 567, KV 005 (JINR) ).

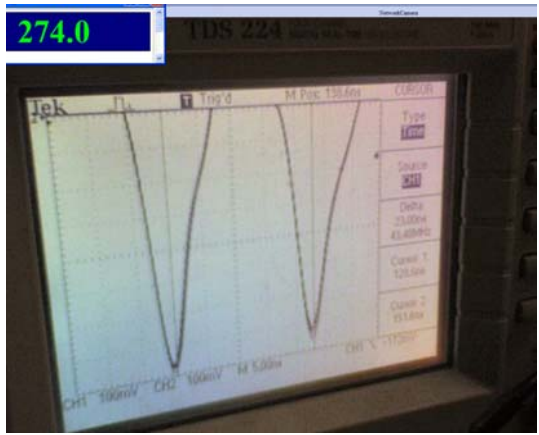

Fig. 4a. A view at the cyclotron operator working place PC. It is shown a beam energy (MeV, $f_{refresh} \sim 1\ c^{-1}$) and "quality" of beam tuning signals from two picUp electrodes located before the separator (oscilloscope picture - via net-camera).

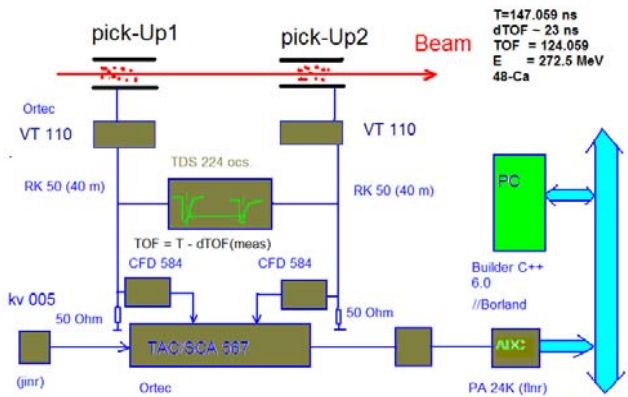

Fig. 4b. TOF U400 system block-diagram

### III. SUMMARY

PC-based experiment parameter monitoring system of the Dubna Gas Filled Recoil separator is designed and successfully applied in experiment aimed to the synthesis of Z=117 element. It allows monitoring the parameters associated with cyclotron beam, detection system of DGFRS and the separator by itself. In some moments it provides fast switching the beam off in the case of detecting the alarm situations. Namely with this system it has became possible to provide long-term irradiations of actinide targets by intense heavy ion beams of $^{48}$Ca.

ACKNOWLEDGMENT

We thank to Drs. V.Utyonkov, I.Shirokovsky and F.Abdullin for assistance in the system testing. This paper is supported in part by the RFBR Grant No.

REFERENCES

[1] A.M.Sukhov, Yu.S.Tsyganov, A.N.Polyakov, "Single event gate rupture on thin gate oxides,"Letters to ECHAYA. 2010/ in print. In Russian/.
[2] Yu.S.Tsyganov, A.N.Polyakov, A.M.Sukhovl, "Control of long term experiment parameters", in the Proc. of Stability and Control Processes Int. Conf. in memory of V.Zubov ; SCR-2005, vol.1 . Saint-Petersburg, Russia. 30.06-01.07, 2005, pp. 234-244 /in Russian/.